\documentclass[12pt]{article}
\usepackage{amsfonts,latexsym,amsmath,latexsym,amssymb,psfig}

\newcommand{\myps}[4]
{\begin{figure}[htb] \label{#4}
    \centerline {\psfig{figure=#1,width=#2 cm}}
    \centerline {\parbox{12cm}{\caption{#3}}}
  \end{figure}}

\newtheorem{definition}{Definition}
\newtheorem{theorem}{Theorem}

\def\C{{\mathbb{C}}}

\newcommand{\sL}[0]{\mathrm{\mathbf{L}}}

\newcommand{\cH}{{\cal H}}

\newcommand{\onemat}[0]{{\mathbf 1}}

\newcommand{\proof}[1]{{\bf Proof.} #1 \hfill $\Box$\vspace{0.5cm}}

\title{{Treating the Independent Set Problem\\ by
$2$D Ising Interactions\\ with Adiabatic Quantum Computing}}

\author{Pawel Wocjan\thanks{e-mail: {\protect\tt
\{wocjan,janzing,eiss\_office\}@ira.uka.de}}, Dominik Janzing, and
Thomas Beth \\ \small Institut {\"u}r Algorithmen und Kognitive
Systeme, Universit{\"a}t Karlsruhe,\\[-1ex] \small Am Fasanengarten 5,
D-76\,131 Karlsruhe, Germany}

\date{February 4, 2003}

\begin{document}

\maketitle

\abstract{We construct a nearest-neighbor Hamiltonian whose ground
states encode the solutions to the NP-complete problem INDEPENDENT SET
in cubic planar graphs. The Hamiltonian can be easily simulated by
Ising interactions between adjacent particles on a $2$D rectangular
lattice. We describe the required pulse sequences. Our methods could
help to implement adiabatic quantum computing by ``physically
reasonable'' Hamiltonians like short-range interactions.
}

\section{Introduction}
Adiabatic quantum computation has been proposed as a general way of
solving computationally hard problems on a quantum computer
\cite{Farhi:01}. Adiabatic quantum algorithms proposed so far work by
applying a time-dependent Hamiltonian
\begin{equation}
H(t) = (1-\frac{t}{T}) H_B + \frac{t}{T} H_P
\end{equation}
that interpolates linearly from an initial Hamiltonian $H_B$ to the
final Hamiltonian $H_P$. The Hamiltonians are chosen such that the
ground states of $H_B$ are easily prepared and the ground states of
the final Hamiltonian $H_P$ encode the solutions to the problem
\cite{Farhi:01}.

The running time of the algorithm is denoted by $T$. If $H(t)$ varies
sufficiently slowly, i.e., $T$ is sufficiently high, then one hopes
that the final state of the quantum computer will be close to the
ground state of the final Hamiltonian $H_P$, so a measurement will
yield a solution to the problem with high probability. The adiabatic
theorem is the justification for this hope. However, it is not clear
whether all necessary conditions for adiabatic evolution are
satisfied. For instance, it is not clear whether the gap between the
ground states and first excited states of $H(t)$ is sufficiently high
for all $t$.

The adiabatic method can only succeed if the Hamiltonian $H(t)$
changes slowly. But how slow is slow enough?  Unfortunately, this
question has proved difficult to analyze in general. Some numerical
evidence suggests the possibility that the adiabatic method might
efficiently solve computationally interesting instances of hard
combinatorial problems, outperforming classical algorithms
\cite{Farhi:01}. Whether adiabatic quantum computing provides a
definite speedup over classical methods for certain problems remains
an interesting open question.

Our objective in this paper is not to explore the computational power
of the adiabatic quantum computing, but rather to investigate how to
implement the adiabatic time evolution starting from ``physically
reasonable'' Hamiltonians like short-range interactions.

A Hamiltonian can be considered as physically reasonable only if it is
``local''. One way to describe locality is a follows. Let $\cH:=\C^2$
denote the Hilbert space of a single qubit and $\cH^{\otimes n}$ the
joint Hilbert space of $n$ qubits. $\sL(\cH^{\otimes s})$ denotes the
set of linear operators from $\cH^{\otimes s}$ to $\cH^{\otimes
s}$. Let $A\in\sL(\cH^{\otimes s})$ be an arbitrary operator and
$S\subseteq\{1,\ldots,n\}$ with $|S|=s$. We denote by
$A[S]\in\sL(\cH^{\otimes n})$ the embedding of the operator $A$ into
the Hilbert space $\cH^{\otimes n}$, i.e., the operator that acts as
$A$ on the qubits specified by $S$.

An operator $H:\cH^{\otimes n}\rightarrow\cH^{\otimes n}$ is called an
$s$-local Hamiltonian if it is expressible in the form
\begin{equation}
H=\sum_j H_j[S_j]\,,
\end{equation}
where each term $H_j\in\sL(\cH^{\otimes |S_j|})$ is a Hermitian
operator acting on a set $S_j$, $|S_j|\le s$.

A Hamiltonian is local if it can be expressed as a sum of terms, where
each term acts on a bounded number of qubits. Indeed, in this case,
the corresponding time evolution can be approximately simulated by a
universal quantum computer \cite{NC:00}.

For a {\em direct physical implementation} of the continuously varying
Hamiltonian $H(t)$ we require a stronger {\em locality condition}.
Physical interactions are usually pair-interactions, unless one
considers effective Hamiltonians. The system Hamiltonian can be thus
decomposed as
\begin{equation}
H = \sum_{k<l} H_{kl} + \sum_k H_k\,,
\end{equation}
$H_{kl}$ is a Hermitian operator acting on the joint Hilbert space of
particle $k$ and $l$ and $H_k$ is the free Hamiltonian of particle
$k$. Furthermore, the interaction strength is decreasing with the
distance. Therefore, we do not want to propose a scheme that relies on
``weak'' interaction terms among distant particles. We thus require
that each particle is coupled to only a few other particles in its
direct neighborhood.

One of the most simple nontrivial examples are the Ising interactions
on a $2$D lattice. Our resource is the Ising Hamiltonian on an
$r\times s$ rectangular lattices, i.e.,
\begin{equation}\label{eq:localHam}
H_{\rm Ising}=\sum_{(k,l)\in L} \sigma_z^{(k)} \sigma_z^{(l)}\,,
\end{equation}
where $L$ are the edges of a rectangular lattice, i.e, a graph of
order $rs$ obtained by placing vertices at the coordinates
$\{(i,j)\mid 0\le i<r, 0\le s<J\}$ with edges joining just the pairs
at unit distance.

Let $L'$ be a subgraph of $L$. We construct a final Hamiltonian
\begin{equation}
\hat{H}_P = \sum_{(k,l)\in L'} w_{kl}\,\sigma_z^{(k)} \sigma_z^{(l)} +
\sum_k\,w_k \sigma_z^{(k)}\quad\mbox{with } w_{kl},w_k\in\mathbb{Z}\,,
\end{equation}
such that its ground states encode the solution to the NP-complete
problem ``independent set''. Clearly, such Hamiltonians satisfy the
locality condition. The aim of our paper is to show how such
Hamiltonians can be constructed using planar orthogonal embeddings of
graphs and how they can be obtained efficiently from the $2$D Ising
model Hamiltonian $H_{\rm Ising}$. Together with the choice of a local
initial Hamiltonian
\begin{equation}
\hat{H}_B = \sum_{k} \sigma_x^{(k)}
\end{equation}
our results allow to simulate efficiently the adiabatic quantum
evolution according to
\[
\hat{H}(t)=(1-\frac{t}{T})\hat{H}_B + \frac{t}{T} \hat{H}_P\,.
\]

\section{Independent set problem}
The idea to consider the INDEPENDENT SET problem is motived by
\cite{KL:02,RAS:02}. The INDEPENDENT SET problem \cite{GJ:79} is
defined as follows:
\begin{itemize}
\item INSTANCE: Graph $G=(V,E)$, positive integer $v\le |V|$.
\item QUESTION: Does $G$ contain an independent set whose cardinality
is at least $v$, i.e., a subset $V'\subseteq V$ such that $|V'|\ge v$
and such that no two vertices in $V'$ are joined by an edge in $E$?
\end{itemize}
The INDEPENDENT SET problem remains NP-complete for cubic planar
graphs \cite{GJS:76}. A graph is called {\em cubic} if all vertices
have degree $3$, i.e., all vertices are connected to exactly three
vertices. A graph is called {\em planar} if it can be drawn in the
plane such that the edges do not intersect. An example of a planar
cubic graph is shown in Figure~1. \myps{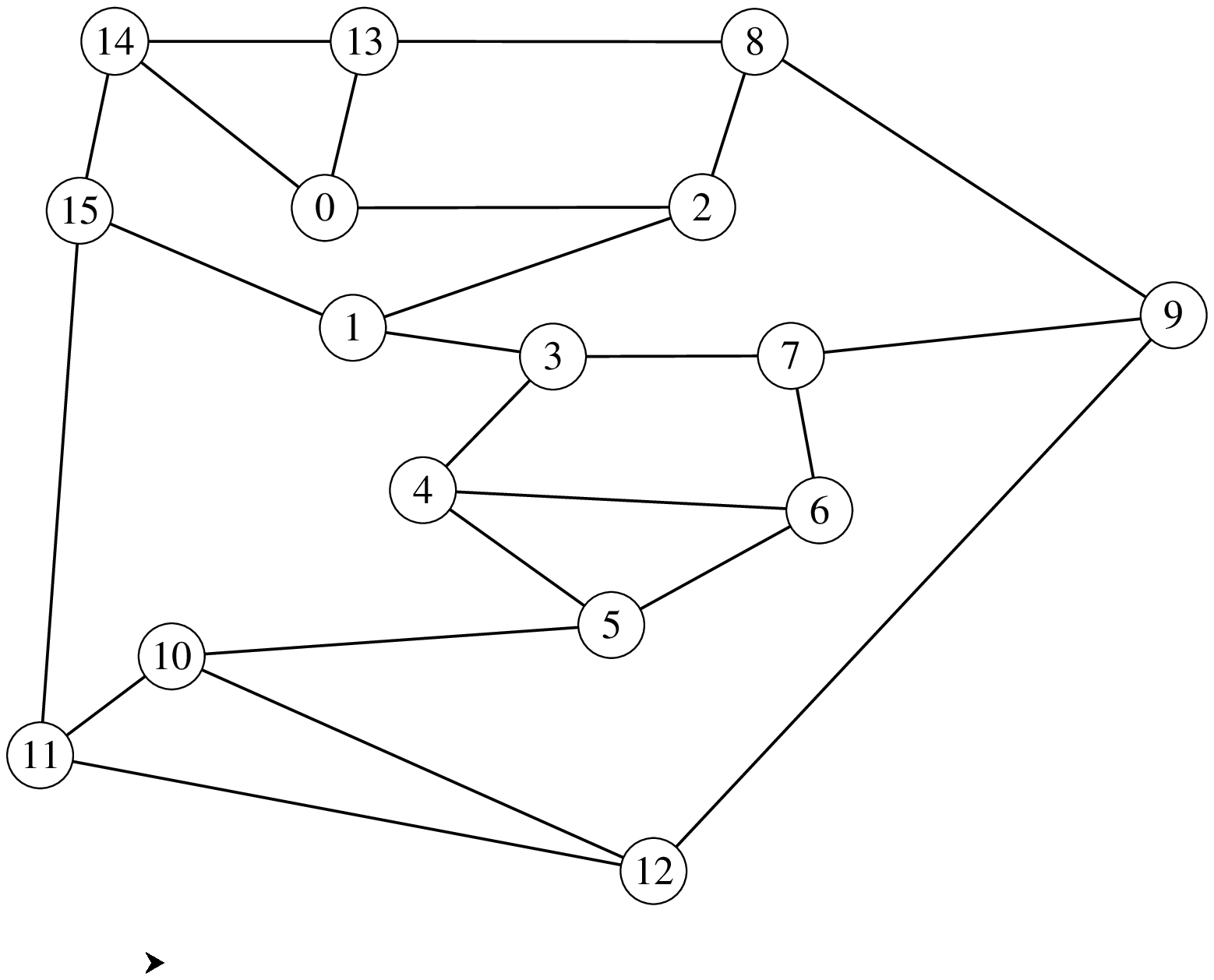}{6}{Planar cubic
graph}{}

We consider a method that determines the maximal cardinality of
independent set of a planar cubic graph. Let us recall how the
solution to the maximum independent set can be encoded in the ground
states of a pair-interaction Hamiltonian $H_P$.
\begin{theorem}[Planar spin glass within a magnetic field]${}$\\
Let $G=(V,E)$ be a cubic planar. Determining the energy of the ground
states of the corresponding Hamiltonian
\begin{equation}
H_P=\sum_{(k,l)\in E} \sigma_z^{(k)}\sigma_z^{(l)} + \sum_{k\in V}
\sigma_z^{(k)}
\end{equation}
is equivalent to determining the maximum cardinality of independent
sets of $G$.
\end{theorem}
\proof{This has been shown in \cite{Barahona:82} (see also
\cite{WB:03}). We include the proof here for completeness. We
associate a variable $X_k\in\{0,1\}$ to each vertex $k\in V$. There is
an independent set whose cardinality is at least $v$ if and only if
there is an assignment to the variables $\{X_k\mid k\in V\}$ such that
\begin{equation}\label{eq:independentSet}
L=\sum_{k\in V} X_k - \sum_{(k,l)\in E} X_k X_l\ge v\,.
\end{equation}
This is seen as follows. If $V'$ is an independent set whose
cardinality is at least $v$, then the assignment $X_k=1$ for $k\in V'$
and $X_k=0$ for $k\in V\setminus V'$ fulfills
inequality~(\ref{eq:independentSet}). 

Now let $X_1,\ldots,X_n$ be an assignment that fulfills
inequality~(\ref{eq:independentSet}). If $V'=\{k\mid X_k=1\}$ is not
an independent set, then we must have $|V'|\ge v+p$, where
$p:=\sum_{(k,l)\in E} X_k X_l>0$ is the ``penalty'' for $V'$ not being
an independent set. Let $(\tilde{k},\tilde{l})\in E$ with
$X_{\tilde{k}}=X_{\tilde{l}}=1$. By removing $\tilde{k}$ from $V'$
(i.e.\ setting $X_{\tilde{k}}:=0$) the cardinality of $V'$ drops by
$1$, while $p$ drops by at least $1$. After repeating this several
times, we end up with an independent set whose cardinality is at least
$v$.

Setting $S_k=2 X_k-1$ for all $k\in V$ and observing that
$|E|=\frac{3}{2}|V|$ for all cubic graphs, we obtain
\begin{equation}
L = -\frac{1}{4}\sum_{k\in V} S_k -
\frac{1}{4}\sum_{(k,l)\in E} S_k S_l + \frac{1}{8} |V|\,.
\end{equation}
For $E=-4 L + \frac{1}{2} |V|$ we see that there exists an independent
set whose cardinality is at least $k$ if and only if there is an
assignment of values to the variables $S_k\in\{-1,1\}$ (corresponding
to the eigenvalues of $\sigma_z$) such that
\begin{equation}
E = \sum_{k\in V} S_k + \sum_{(k,l)\in E} S_k S_l\le
\frac{1}{2}|V| - 4 v\,.
\end{equation}
Now it is clear that determining the minimal energy $E$ is equivalent
to determining the maximal cardinality $v$ of independent sets of
$G$.}

In adiabatic quantum computing the initial Hamiltonian is chosen as
\begin{equation}
H_B=\sum_{k\in V} \sigma_x^{(k)}
\end{equation}
and the time-dependent Hamiltonian as
\begin{equation}
H(t)=(1-\frac{t}{T}) H_B + \frac{t}{T} H_P\,.
\end{equation}
This Hamiltonian $H_P$ does not necessarily satisfy the locality
conditions.

\section{``Planar orthogonal'' Hamiltonians}
Due to the lattice structure of our resource Hamiltonian we need to
embed our graph into this structure. This can be done using {\em
planar orthogonal embeddings} of graphs \cite{KW:01}. This idea is
inspired by \cite{KL:02,RAS:02}. We shall be concerned with embedding
graphs into a $2$D rectangular lattice.

\begin{definition}[Planar orthogonal embedding]${}$\\
A planar orthogonal embedding $\Gamma$ of a graph $G=(V,E)$ is a
mapping that
\begin{itemize}
\item maps vertices $k\in V$ to lattice points $\Gamma(k)$ and 
\item edges $(k,l)\in E$ to paths in the lattice such that the images of
their endpoints $\Gamma(k)$ and $\Gamma(l)$ are connected and such
that the paths do not share any vertices (besides the endpoints).
\end{itemize}
\end{definition}
Note that the map inserts ``dummy vertices'' if necessary to create
the paths connecting the vertices $\Gamma_k$ and $\Gamma_l$. A planar
orthogonal embedding is shown in Figure~2.
\myps{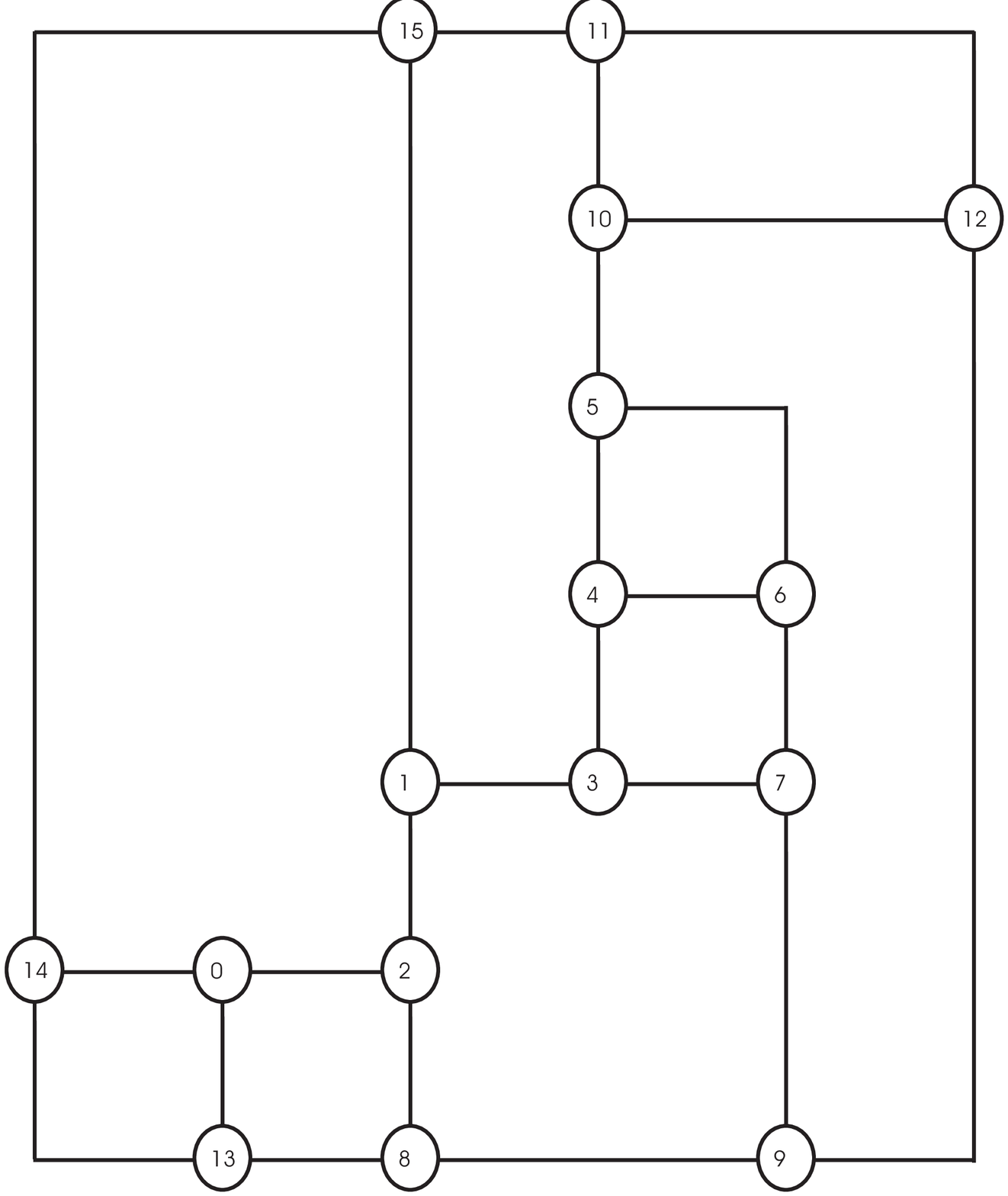}{6}{Planar orthogonal embedding of the graph in
Fig.~1}{}

Every planar graph with maximum degree $3$ admits a planar orthogonal
embedding on an $\lfloor n/2\rfloor \times \lfloor n/2\rfloor$.  The
algorithm presented in \cite{Kant:96} computes efficiently such planar
orthogonal embeddings of graphs. We used AGD (Libary of Algorithms for
Graph Drawing) to compute the embedding \cite{AGD}.

In the proposal of \cite{KL:02} the Hamiltonian $H_P$ is considered.
The planar orthogonal embedding gives a regular wiring among the
qubits. This means that the couplings are not spatially local. In
contrast, we need a Hamiltonian $\hat{H}_P$ that contains only
nearest-neighbor interactions. This is necessary that it can be
simulated by $H_{\rm Ising}$.  The idea is to use the dummy vertices
as wires that propagate the state of a (real) vertex spin to the
neighborhood of another vertex.  This can be achieved by constructing
a path of adjacent dummy vertices, each interacting with its neighbor
by a strong ferromagnetic coupling. Furthermore, the first dummy at
one end of this ``dummy path'' is strongly ferromagnetically coupled
to a vertex and the last dummy at the other end is in the neighborhood
of another real vertex, coupled to it via a usual antiferromagnetic
interaction. The interaction strength is chosen in such a way that it
is always energetically better when all dummies have the same state as
the real vertex to which they are connected to than to have a mismatch
along the ``ferromagnetic path''.

Formally, this construction is as follows:
\begin{itemize}
\item The dummy vertices have no local $\sigma_z$ term.
\item The vertices $\Gamma(k)$ have $\sigma_z$ as local Hamiltonians.
\item Let $(k,l)\in E$ be an edge of $G$. 

If $\Gamma_k$ and $\Gamma_l$ are adjacent, then the coupling between
$\Gamma_k$ and $\Gamma_l$ is chosen to be antiferromagnetic, i.e.,
$\sigma_z\otimes\sigma_z$.

Otherwise there are $m$ dummy vertices $v_1,\ldots,v_m$ such that the
path $(\Gamma_k,v_1,\ldots,v_m,\Gamma_l)$ connects the vertices
$\Gamma_k$ and $\Gamma_l$. The couplings between $\Gamma_k$ and $v_1$
and $v_i$ and $v_{i+1}$ for $i=1,\ldots,m-1$ are chosen to be
ferromagnetic with coupling strength $c$, i.e.,
$-c\,\sigma_z\otimes\sigma_z$. The coupling between $v_m$ and
$\Gamma_l$ is chosen to be antiferromagnetic, i.e.,
$\sigma_z\otimes\sigma_z$.
\end{itemize}
The corresponding ``planar orthogonal Hamiltonian'' is shown in
Figure~3. The filled circles correspond to dummy vertices that do not
have any local Hamiltonian. The circles with indices correspond to the
original vertices of $G$. They have $\sigma_z$ as local
Hamiltonians. The thin lines correspond antiferromagnetic interactions
and the thick lines to ferromagnetic interactions.
\myps{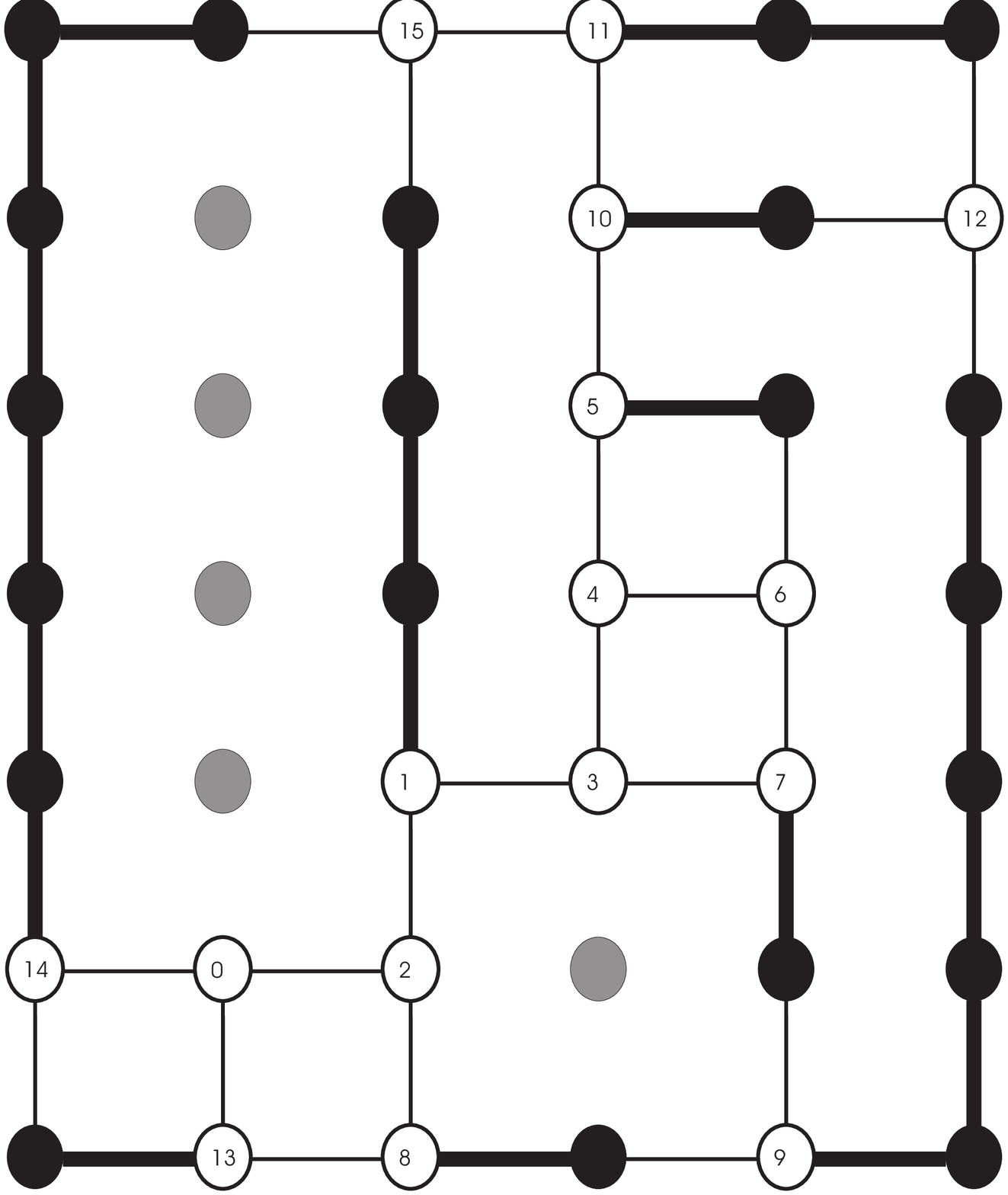}{6}{Hamiltonian corresponding to the planar
orthogonal embedding in Fig.~2}{} The idea behind this construction is
that there is a direct one-to-one correspondence between the ground
states of $H_P$ and $\hat{H}_P$. The same is true for the first
excited states. This can be seen as follows:

Let $(k,l)\in E$ be an edge of $G$ and
$(\Gamma_1,v_1,\ldots,v_m,\Gamma_l)$ be the path on the lattice
connecting $\Gamma_k$ and $\Gamma_l$. The variables
$S_{\Gamma_k},S_1,\ldots,S_m \in\{0,1\}$ indicate whether the
corresponding qubit is spin up or spin down.

A ground state satisfies the condition that $S_1,\ldots,S_m$ are all
equal to $S_{\Gamma_k}$. To show this we define the number of
mismatches along the path to be the number of occurrences of
$S_{\Gamma_k}\neq S_1$, $S_i\neq S_{i+1}$ for $i=1,\ldots,m-1$. This
number is denoted by $\delta$.

Then the minimal possible energy (due to the couplings along the path)
is
\begin{equation}
c(-m+\delta)-1\,.
\end{equation}
If we remove the mismatches (by setting $S_i:=S_{\Gamma_k}$ for
$i=1,\ldots,m$) then the maximal possible energy is
\begin{equation}
-c m+1\,.
\end{equation}
By choosing $c=3$ minimal energy can be achieved only if the states of
all dummy vertices are equal to the state of the qubit corresponding
to $\Gamma_k$.

For adiabatic quantum computing it is important that the gap between
the ground and first excited states of the Hamiltonian at all times is
sufficiently large. We show that the modification of $H_P$ to
$\hat{H}_P$ does not decrease this gap.

The gap between the ground and the first excited states of $H_P$ is
smaller or equal to $8$. This is seen as follows. Let
$S_1,\ldots,S_n\in\{-1,+1\}$ be an assignment corresponding to a
ground state of $H_P$. Pick any vertex $k$ and let $l_1,l_2,l_3$ be at
the three vertices connected to $k$. By flipping $S_k$ the energy can
increase by at most $8$ because the relevant Hamiltonian is
\[
\sigma_z^{(k)} + \sum_{i=1}^3 \sigma_z^{(k)} \sigma_z^{(l_i)}\,.
\]
By choosing $c=9$ it is seen that the first excited states of $\hat{H}_P$
satisfy the condition that the states all of dummy vertices are equal
to the vertex of $\Gamma_k$.

\section{Simulating ``planar orthogonal''\\ Hamiltonians}
To implement the time-evolution according to the Hamiltonian
$\hat{H}_P$ we make use of the concepts of simulating Hamiltonians
that has been used in nuclear magnetic resonance for a long time
\cite{EBW:87}. These techniques rely on the so-called average
Hamiltonian approach. The idea is to conjugate the natural time
evolution by unitary control operations $u_j$, i.e., the total
evolution is
\[
u_k\exp(-iHt_k) u_k^\dagger \dots  
u_2^\dagger \exp(-iHt_2) u_2^\dagger u_1 \exp(-iHt_1) u_1^\dagger\,,
\]
where the system evolves in an undisturbed way during periods of length
$t_1,t_2,\dots,t_k$.
If these periods are short compared to the time scale of the natural 
evolution, the total dynamics is approximatively the same as
the evolution according to the average Hamiltonian
\[
\overline{H}:= \sum_j \frac{t_j}{t} u_j H u_j^\dagger
\]
with $t:=\sum_j t_j$.
Usually, the control operations on $n$ particles 
are assumed to be of the form
\[
u:=v_1\otimes v_2 \otimes \dots \otimes v_n
\]
where $v_j$ is a unitary acting on particle $j$.  The design of
simulation schemes for Hamiltonians with $n$ particles interacting via
pair-interactions leads to non-trivial combinatorial problems (e.g.\
\cite{LCY:00,DNBT:01,WJB:02,JWB:02}). An experimental proposal for
simulating dynamics in optical systems is presented in
\cite{JVDZC:02}.

Starting from the Ising Hamiltonian $H_{\rm Ising}$, we can implement
the Hamiltonian $\hat{H}_P$ with time overhead (slow-down) $2c+1$ and
$16$ time steps by interspersing the time evolution according to
$H_{\rm Ising}$ by local operations in $X\otimes X\otimes\cdots\otimes
X$, where $X=\{\onemat,\sigma_x\}$.

Following the ideas of \cite{LCY:00,WJB:02} we construct a selective
decoupling scheme based on Hadamard matrices. Due to the special form
of $H_{\rm Ising}$ it is sufficient to use the Hadamard matrix of size
$4$ only.

Our scheme consists of $4$ subroutines that implement the following 
couplings of $\hat{H}_P$:
\begin{enumerate}
\item horizontal $\sigma_z\otimes\sigma_z$, 
\item vertical $\sigma_z\otimes\sigma_z$, 
\item horizontal $-c\,\sigma_z\otimes\sigma_z$, and
\item vertical $-c\,\sigma_z\otimes\sigma_z$
\end{enumerate}

The indices $i,j$ enumerate the rows and the columns of the lattice.
We denote the columns of the Hadamard matrix of size $4$ 
\[
W:=\left(
\begin{array}{rrrr}
1 &  1 &  1 &  1 \\
1 & -1 &  1 & -1 \\
1 &  1 & -1 & -1 \\
1 & -1 & -1 &  1
\end{array}
\right)
\]
by $W(0,0), W(0,1), W(1,0)$ and $W(1,1)$. 

Let $v=(v_1,v_2,v_3,v_4)\in\{-1,1\}^4$ be a column vector. We use the
abbreviation
\begin{quote}
``apply $v$ at $(i,j)$''
\end{quote}
to denote the following control sequence with $4$ equally long time
steps: at the beginning and the end of the $s$th time step we apply
$\sigma_x$ on the qubit at position $(i,j)$ if $v_s=-1$ and do nothing
if $v_s=1$, where time step $s$ runs from $1$ to $4$.

Let $v,v'\in\{-1,1\}^4$. One easily checks that applying $v$ and $v'$
at adjacent lattice points changes $\sigma_z\otimes\sigma_z$ to
$\langle v,v'\rangle\,\sigma_z\otimes\sigma_z$, where $\langle
v,v'\rangle$ denotes the inner product of $v$ and $v'$. This is the
key observation for constructing the selective decoupling scheme.

In the first and second subroutines the length of the $4$ time steps
is chosen to $1/4$. Let us consider the first subroutine. The vertical
couplings are automatically removed if we apply in rows with even
indices only $W(0,0)$ and $W(1,0)$ and in rows with odd indices
$W(1,0)$ and $W(1,1)$. The choice between $W(a,0)$ and $W(a,1)$
depends on which horizontal interactions should remain or be switched
off. Explicitly, this choice is as follows. Choose $W(a,0)$ for the
leftmost spin.  If the interaction between the spins $(j-1)$ and $j$
should remain, then apply the same $W(a,b)$ to $j$ as to
$(j-1)$. Otherwise (i.e.\ the coupling should be switched off) apply
the second possible $W(a,b')$ to $j$.

The second subroutine is obtained from the first subroutine by
exchanging the roles of rows and columns of the lattice.

In the third and fourth subroutines the length of the $4$ time steps
is chosen to $c/4$. The third subroutine is obtained from the first
subroutine by apply $(-1)^j v$ instead of $v$ to the spin
$j$. Finally, the fourth subroutine is obtained from the third
subroutine by exchanging the roles rows and columns of the lattice.

\section*{Acknowledgments}
This work was supported by grants of the BMBF project MARQUIS
01/BB01B. 

We would like to thank Helge Ros{\'e}, Torsten Asselmeyer, and Andreas
Schramm for interesting discussions that led us to consider this
problem. Therese Biedl, Thomas Decker and Khoder Elzein provided
interesting discussions on orthogonal graph drawing and installed the
graph drawing software.

\newcommand{\etalchar}[1]{$^{#1}$}

\end{document}